# Data fusion of multivariate time series: Application to noisy 12-lead ECG signals

Chen Diao[1,2*], Bin Wang[1,2]

1. School of Electrical Engineering, Northwest Minzu University, Lanzhou 730030, China
2. Key Laboratory of National Language Intelligent Processing, Gansu Province, Northwest Minzu University, Lanzhou, 730030, China
Correspondence should be addressed to Chen Diao; diaoc80@163.com

12-lead ECG signals fusion is crucial for further ECG signal processing. In this paper, a novel fusion data algorithm is proposed. In the method, 12-lead ECG signals are appropriately converted to a single-lead physiological signal via the idea of the local weighted linear prediction algorithm. For effectively inheriting the quality characteristics of the 12-lead ECG signals, the fuzzy inference system is rationally designed to estimate the weighted coefficient in our algorithm. Experimental results indicate that the algorithm can obtain desirable results on synthetic ECG signals, noisy ECG signals and realistic ECG signals.

## 1. Introduction

ECG records the physiological information of cardiac activity by using some electrodes on different positions of the body. Therefore, ECG recordings have been widely applied to clinic diagnosis and clinical monitor. Nevertheless, the ECG recordings gathered in clinical settings are often contaminated by noise and artifacts. On account of the overlapping frequency bands and similar morphologies in noise and ECG signal [1, 2], the characteristics of ECG signal are usually distorted and result in the false alarms (FA) in intensive care unit (ICU) and imprecise measurement of ECG signal [2]. Thus, ECG signal quality assessment is a significant element for further ECG signal processing.

The quality condition of ECG signal is related to the accuracy of ECG analysis. Thus the ECG quality assessment algorithms have been successfully developed [3, 4].

The key of these algorithms is to extract the characteristics of the ECG signal appropriately. For instance, as time domain approach, Moody *et al.* adopted Karhumen-Loeve basis functions to represent the QRS complex and employed the residual error of the reconstructed QRS complex to estimate the instantaneous noise of the original signal [3]. On the other hand, as frequency domain approach, via a long-term ECG recording from coronary care unit, the frequency content and the number of times the ECG exceeded a preset limit were employed to analyze the ECG quality [4].

PhysioNet, a resource for biomedical research sponsored by the National Institutes of Health (NIH), solicited algorithms in 2011 to effectively assess the ECG quality via mobile telephones [5]. Subsequently, in order to estimate signal quality accurately, many ECG quality assessment algorithms were proposed [6-15]. Among the majority of the proposed algorithms, the time-domain characteristics or the frequency-domain characteristics are used alone for ECG quality assessment. Jekova *et al.* employed time-frequency-domain characteristics for assessing the ECG quality, yielding a sensitivity of 81.8% and a specificity of 97.8% [9]. Clifford *et al.* presented some signal quality indices (SQIs) and these characteristics also involved both time-domain and frequency-domain, which can partly reflect the condition of the ECG quality [10]. In [11], Li *et al.* developed four novel signal quality indices which enriched the SQIs in some way. The SQIs can provide a great deal of physiological information, thus those quality indices have been extensively used for ECG quality assessment [2, 11, 12].

Apparently extracting the ECG quality characteristics reasonably is significant for the quality assessment algorithm. Nonetheless, the conventional methods can only be performed on a single-lead ECG signal for assessing the ECG quality. Once facing 12-lead ECG signals, the signal quality of each lead must be assessed individually, thereby the computational efficiency is brought down.

In order to reduce the amounts of pending ECG data properly, Chen *et al.* employed Dower transform to convert 12-lead ECG signals to 3-lead vectorcardiogram (VCG), which are properly analyzed by multiscale recurrence analysis in each scale [16]. Wavelet analysis is an effective algorithm for handling nonlinear and nonstationary signal. Nevertheless, VCG signals are decomposed into a series of multiple wavelet scales and this also increases the amount of pending data observably. In other words, in [16], the application of multiscale recurrence analysis virtually weakened the original ideal of the Dower transform.

Principal component analysis (PCA) is a significant technique of signal processing derived from applied linear algebra. As a simple and non-parametric method, PCA has been applied to ECG analysis successfully [17]. The purpose of PCA is to map high-dimensional data to low-dimensional spaces, thereby realizing data compression. Nevertheless, in PCA, the most interesting variances of data set are associated with the first $k$ principal components and other less important components, which can help revealing the dynamics information of high-dimensional data further but are discarded [18]. Hence, the loss of data information is inevitable to some degree when using PCA. The time series of ECG signals are essentially both nonlinear and nonstationary. In order to comprehensively analyze ECG signals, the phase space reconstruction theory has also been applied [19, 20]. According to this theory, the ECG signals are reconstructed in high-dimensional spaces with more complete characteristic information of the original signals. In this regard, the approaches based on phase space reconstruction are relatively preferable to PCA for 12-lead ECG signals fusion.

12-lead ECG signal contains abundant heart state information and can fully describe the electrical activity of the heart. Because the electrodes of the twelve leads are placed in the different position of the body, the signal amplitudes are different between all lead ECG signals. It means that there are differences for the time-domain characteristics of the twelve leads and the signal qualities of all lead signals ought to be assessed respectively. In order to analyze the quality of the 12-lead ECG signals in full directions, many characteristic parameters of each lead of ECG signals, e.g. SQIs, need to be calculated. Facing the quality assessment of 12-lead ECG signals, conventional algorithms are virtually inefficient in relative terms. Thus how to comprehensively assess the quality of 12-lead ECG signals and effectively reduce the computational complexity are very critical. The key solution lies in converting the 12-lead ECG signals into a single-lead physiological signal. It means that the quality characteristics of original signals are inherited in the single-lead signal as much as possible.

The main objective of this study is to develop a novel fusion data algorithm (NFDA) for 12-lead ECG signals that can adequately integrate the qualitative characteristics of 12-lead ECG signals into a single-lead signal. Previous studies implied that ECG signals possess the nonlinear and nonstationary characteristics, which is the chaotic signal [19]. Currently, the analysis of deterministic chaos is an active field. In many branches of the research, chaotic time series prediction is a fundamental issue of chaos theory [21, 22]. Previous research has shown that the local weighted linear prediction algorithm (LWLPA) of chaotic time series has better performance than the global chaotic time series prediction algorithm [23]. In LWLPA, chaotic time series are reconstructed in high dimensional space based on the theory of phase space reconstruction and the whole evolutionary information of original signal can be fully reflected in the high dimensional space theoretically. Some neighboring states on evolutionary trajectories are used to predict their future states by a linear prediction model. Inspired by LWLPA, the idea of the algorithm is appropriately utilized in the NFDA algorithm, which can transform the 12-lead ECG into a single-lead signal.

The main contributions of the paper are twofold. (1) The idea of LWLPA is first applied to fuse 12-lead ECG signals. (2) In NFDA, two FISs are designed to properly calculate the weighted coefficient of each lead ECG signal.

The outline of the rest of this paper is as follows. In Section 2, LWLPA is briefly discussed as preliminary. Section 3 introduces NFDA which is based on LWLPA. The performance of NFDA is evaluated by synthetic ECG signals and realistic ECG signals in Section 4. Section 5 contains the conclusion.

## 2. The Local Weighted Linear Prediction Algorithm

In the course of the signal quality assessment of 12-lead ECG signals, compressing the pending ECG data is an efficient solution to further improve the efficiency of the assessment algorithms. Since the cardiac signals reveal the possibility of deterministic chaos, here the LWLPA algorithm as an important prediction method of chaotic time series is used to fuse 12-lead ECG signals. In this section, as a preliminary, we will briefly review the algorithm, which is closely related to NFDA.

Due to the butterfly effect of chaotic systems, the evolutionary tendency of chaotic systems cannot be predicted in a relatively long time. However, there still exists of predictability because of the linear correlation of motions in short period for chaotic systems. Takens [24] proved that if the embedding dimension and delay time can be chosen appropriately, the regular evolutionary trajectory of chaotic systems could be completely reconstructed and revealed in an $m$-dimensional space.

For a chaotic time series $\{x(i), i=1,2,\cdots,N\}$, according to the phase space reconstruction theorem, the dynamics characters of the chaotic time series can be well reflected in $m$-dimensional space via the vectors $X_t = (x(t), x(t+\tau), \cdots, x(t+(m-1)\tau)) \in R^m$, $t=1,2,\cdots,N-(m-1)\tau$, where $m$ is the embedding dimension and $\tau$ the delay time. Suppose that $X_k$ is the current state point of the chaotic system and its next step is $X_{k+1}$, the future state of the system needs to be predicted. In LWLPA, the neighboring states $\{X_{ki}, i=1,2,\cdots,n\}$ of the current state point $X_k$ need to be chosen from the reconstructed trajectory. With the neighborhood $X_{ki}$ and the linear prediction model, the future state $X_{k+1}$ can be approximatively estimated.

The steps of the LWLPA algorithm are listed as follow.

1) Select proper parameters of embedding dimension $m$ and delay time $\tau$, and the vectors are constructed as

$$X_t = (x(t), x(t+\tau), \cdots, x(t+(m-1)\tau)) \in R^m, \qquad (1)$$

where $t=1,2,\cdots,N-(m-1)\tau$.

2) Calculate the Euclidean distances between $X_k$ and other neighboring vectors $\{X_{ki}, i=1,2,\cdots,n\}$

$$d_{ki} = \|X_{ki} - X_k\|_2, \qquad (2)$$

and the weighted coefficient of each neighboring state vector $X_{ki}$ is computed as

$$\omega_{ki} = \frac{\exp[-\lambda(d_{ki} - d_{\min})]}{\sum_{i=1}^{n} \exp[-\lambda(d_{ki} - d_{\min})]}, \qquad (3)$$

where $d_{min}$ is the minimum distance in $\{d_{ki}, i = 1, 2, \cdots, n\}$ and $\lambda$ is the regularization parameter being usually set to 1.

3) Estimate the future state $X_{k+1}$ via linear prediction model

$$\begin{bmatrix} X_{k1+1} \\ X_{k2+1} \\ \vdots \\ X_{kn+1} \end{bmatrix} = \begin{bmatrix} e & X_{k1} \\ e & X_{k2} \\ \vdots & \vdots \\ e & X_{kn} \end{bmatrix} \begin{bmatrix} a \\ b \end{bmatrix}, \qquad (4)$$

where $\{X_{ki+1}, i = 1, 2, \cdots, n\}$ is the future state of neighboring vectors $\{X_{ki}, i = 1, 2, \cdots, n\}$ and $e = [1, \cdots, 1]_m^T$.

4) Compute the linear fitting parameters $a$ and $b$ by the least square equation

$$g(a,b) = \min_{a,b \in R} \{\sum_{i=1}^{n} \omega_{ki}(X_{ki+1} - ae - bX_{ki})^2\}. \qquad (5)$$

5) Calculate the one-step prediction state $X_{ki+1}$ of neighboring vectors and the future version $\hat{X}_{k+1}$ of the current state $X_k$ can be estimated as

$$\hat{X}_{k+1} = \sum_{i=1}^{n} \omega_{ki} X_{ki+1}. \qquad (6)$$

6) The predicted value of chaotic time series can be extracted from the state vectors $\hat{X}_{k+1}$.

In LWLPA, the neighboring vectors are used to predict the future state of the chaotic time series via the linear prediction model effectively. Previous researches have implied that the method can exceed the global chaotic time series prediction algorithm in performance, thereby it is widely applied in predicting the chaotic time series. From the data compression perspective, neighboring points in the reconstructed space are converted to an estimated state point. Inspired by the method, the NFDA algorithm will be devised in the next section.

## 3. The Novel Fusion Data Algorithm

As the most important section of this paper, the basic idea of the NFDA algorithm will be briefly introduced in the subsection 3.1. In NFDA, the significance of weighted coefficients and how to obtain these appropriate parameters will be discussed in the subsection 3.2.

### 3.1. Basic Idea of Novel Fusion Data Algorithm

How to significantly improve the efficiency of ECG quality assessment algorithm is a realistic issue. It will facilitate solving this problem if the pending 12-lead ECG signals are effectively compressed. Evidently the LWLPA algorithm can successfully meet the requirement of the problem above.

To illustrate the basic idea of our algorithm, an example will be given. According to the phase space reconstruction theorem, consider the two reconstructed phase trajectories $L_1$ and $L_2$, shown in Fig. 1. In this example, suppose that the trajectory $L_F$ is the fused result of the trajectories $L_1$ and $L_2$. Furthermore, the state point $X_F$ on the trajectory $L_F$ should satisfy the linear prediction model $X_F(p+1) = ae + bX_F(p)$, where $e = [1,\cdots,1]_m^T$. Here, how to obtain the parameters $a$ and $b$ in the linear prediction model is a critical problem.

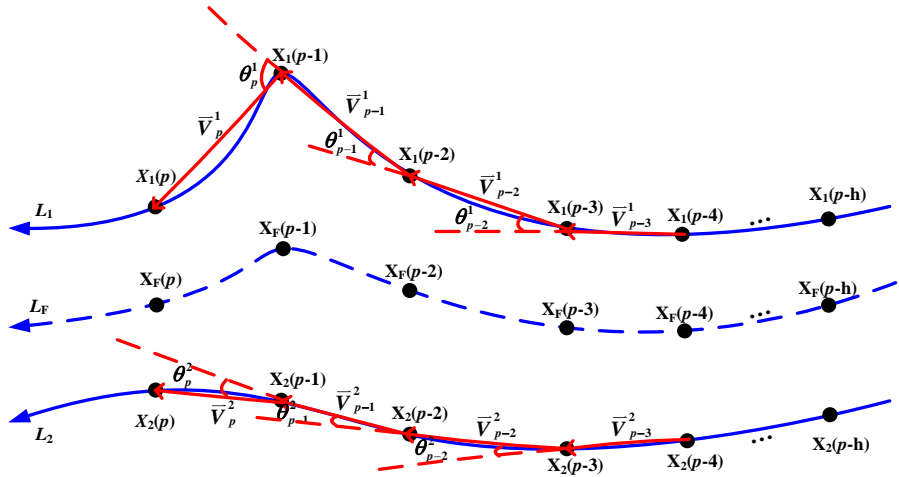

**Figure 1.** Basic idea of NFDA.

From the point of LWLPA, in Fig. 1, the vectors $X_1(p)$ and $X_2(p)$ can be considered as the neighboring vectors of the current state $X_F(p)$. With the two vectors being employed, we can calculate the parameters $a$ and $b$ by (5). The equation can be described as follows

$$g(a,b) = \min_{a,b \in R}\{\sum_{s=1}^{2} \omega_s [X_s(p+1) - ae - bX_s(p)]^2\}, \tag{7}$$

where $\omega_s$ is the weighted coefficient, which reflects the degree of impact from the state point to the fusion result. Then the fused state can be calculated as

$$X_F(p+1) = ae + bX_F(p). \tag{8}$$

The fused trajectory $L_F$ will be employed for the original signal quality assessment. It implies that to some extent, the characteristic information of original signal ought to be fused in the trajectory $L_F$. Here, how to effectively inherit the characteristic information to fused result is a key for NFDA. For solving the problem, the weighted coefficients of state points on evolutionary trajectory should be appropriately estimated.

## 3.2. Weighted Coefficient Design for NFDA

The motivation of weighted coefficients design originated from the idea that, in order to comprehensively analyze the original signal quality, the quality characteristics of original signal should be adequately retained in the final fused result. The values for weighted coefficient of state points are closely related to the final fused result. In Fig. 1, there is an evident amplitude of the point $X_1(p\text{-}1)$ on the trajectory $L_1$ and the time-domain characteristic should be well inherited in the fused trajectory $L_F$. Thus, a larger value for weighted coefficient $\omega_1$ of the state $X_1(p\text{-}1)$ should be chosen, which will further enhance the impact on the final result. Thus how to select a feasible weighted coefficient is the key issue. Here, the Euclidean distance of two neighboring points and the angle between two neighboring vectors on evolutionary trajectory $L$ are used to estimate the weighted coefficients.

In Fig. 1, $X_1(p)$ and $X_2(p)$ are the $p$-step state points on the evolutionary trajectories $L_1$ and $L_2$, respectively. The vector can be easily calculated through two adjacent state points, e.g., for the evolutionary trajectories $L_1$, $\vec{V}_p^1 = X_1(p) - X_1(p-1)$, and then the angle $\theta_p^1$ can be obtained by the two neighboring vectors $\vec{V}_p^1$ and $\vec{V}_{p-1}^1$. The modulus of vector $\vec{V}_p^1$ is the Euclidean distance between $X_1(p)$ and $X_1(p\text{-}1)$.

For the *l*th evolutionary trajectory $L_l$, the modulus of vector $\vec{V}^l$ and the angle $\theta^l$ can well reflect the evolutionary trend of the trajectory. In Fig. 1, for the trajectory $L_1$, the values of the modulus of $\vec{V}_p^1$, $\vec{V}_{p-1}^1$ and the vector angles $\theta_p^1$, $\theta_{p-1}^1$ are larger than the values of the trajectory $L_2$ at the same step. The characteristics of evolutionary trajectory can be described by them objectively. According to the idea, the change of evolutionary trajectory is positively relational to the values of the two parameters. Based on the relationship, the weighted coefficient of the data point can be estimated approximately.

As an important application of fuzzy logic and fuzzy sets theory [25], fuzzy inference system (FIS) has been successfully applied in decision support tools and other subjects. FIS is useful for dealing with linguistic concepts, which can achieve nonlinear mappings between inputs and outputs. Thus, FIS could be well applicable to estimate the weighted coefficient of the data point.

In this subsection, the two simple $FIS_d$ and $FIS_\alpha$ will be devised. $FIS_d$ is composed of nine fuzzy rules. The input variables of $FIS_d$ are the distance $D$ and the change rate $D_r$ of the distance, respectively. The output variable of $FIS_d$ is $O_d$.

The two parameters $D$ and $D_r$ can be calculated as

$$D(p) = \|X(p) - X(p-1)\|_2, \tag{9}$$

$$D_r(p) = |D(p) - D(p-1)|, \tag{10}$$

where $D(p)$ and $D_r(p)$ are the Euclidean distance and the change rate of $X(p)$ at *p*-step, respectively. $FIS_d$ is applied to estimate the weighted coefficient $\omega_d$ via the parameters $D$ and $D_r$.

Similarly, $FIS_\alpha$ consists of two inputs, one output and fifteen fuzzy rules. The two input variables are the angle of cosine $\alpha$ and the change rate $\alpha_r$ of the cosine, which can be obtained as

$$\alpha(p) = \frac{V(p) \cdot V(p-1)}{|V(p)||V(p-1)|}, \tag{11}$$

$$\alpha_r(p) = |\alpha(p) - \alpha(p-1)|, \tag{12}$$

where $\alpha(p)$ is the angle of cosine between two neighboring vectors $V(p)$ and $V(p\text{-}1)$; and $\alpha_r(p)$ the change rate of the cosine at $p$-step. Here, the output variable of $\text{FIS}_\alpha$ is defined as $O_\alpha$. From the angle and its change rate perspective, the weighted coefficient $\omega_\alpha$ could be properly estimated by $\text{FIS}_\alpha$.

Evidently, a significant change of evolutionary trajectory will lead to a larger values of $D$ and $D_r$. In other words, the values of the two parameters $D$ and $D_r$ are positively relational to the change rate of evolutionary trajectory. What is interesting is that there is a negative correlation between the two parameters, $\alpha$ and $\alpha_r$, and the change of evolutionary trajectory. With the four parameters being employed, $\text{FIS}_d$ and $\text{FIS}_\alpha$ can be successfully designed.

In the study, the universe of input $\alpha$ is set to $[-1,1]$ and the other variables of two FISs will be set to $[0,1]$ uniformly. Here, the universes of all variables of the two FISs are divided into several fuzzy sets, shown in Fig. 2. According to the aforementioned relationship, the rules of the two FISs could be devised, which are summarized in Table 1 and Table 2, respectively.

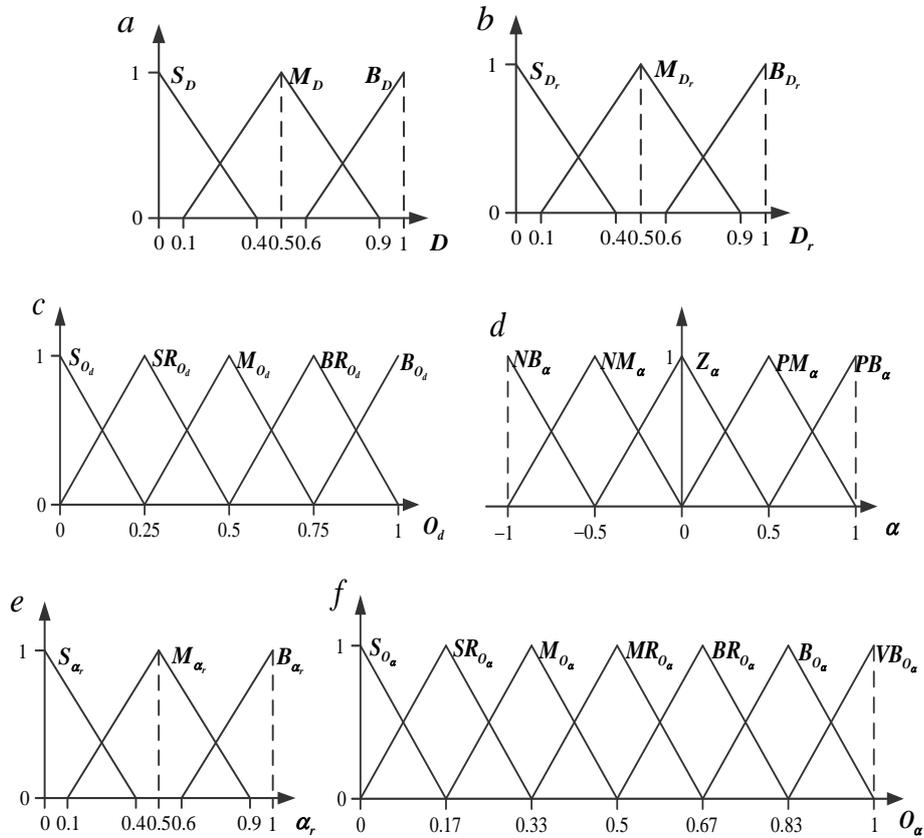

**Figure 2.** Partitioning results of variables with triangular membership function. (a)~(f) are partitioning results of $D$, $D_r$, $O_d$, $\alpha$, $\alpha_r$ and $O_\alpha$, respectively.

**Table 1:** Inference rules of $FIS_d$

| | | D | | |
|---|---|---|---|---|
| | | $S_D$ | $M_D$ | $B_D$ |
| $D_r$ | $S_{D_r}$ | $S_{O_d}$ | $SR_{O_d}$ | $M_{O_d}$ |
| | $M_{D_r}$ | $SR_{O_d}$ | $M_{O_d}$ | $BR_{O_d}$ |
| | $B_{D_r}$ | $M_{O_d}$ | $BR_{O_d}$ | $B_{O_d}$ |

**Table 2:** Inference rules of $FIS_\alpha$

| | | $\alpha$ | | | | |
|---|---|---|---|---|---|---|
| | | $NB_\alpha$ | $NM_\alpha$ | $Z_\alpha$ | $PM_\alpha$ | $PB_\alpha$ |
| $\alpha_r$ | $S_{\alpha_r}$ | $VB_{O_\alpha}$ | $B_{O_\alpha}$ | $BR_{O_\alpha}$ | $MR_{O_\alpha}$ | $M_{O_\alpha}$ |
| | $M_{\alpha_r}$ | $B_{O_\alpha}$ | $BR_{O_\alpha}$ | $MR_{O_\alpha}$ | $M_{O_\alpha}$ | $SR_{O_\alpha}$ |
| | $B_{\alpha_r}$ | $BR_{O_\alpha}$ | $MR_{O_\alpha}$ | $M_{O_\alpha}$ | $SR_{O_\alpha}$ | $S_{O_\alpha}$ |

With FIS being employed, the weighted coefficient of $\omega$ is calculated as

$$\omega = \frac{\sum_{q=1}^{h} \beta(q) y(q)}{\sum_{q=1}^{h} \beta(q)}, \tag{13}$$

where $h$ is the number of rules of FIS, $y(q)$ is the output of the $q$th rule and $\beta(q)$ the degree of activation for the $q$th rule. Based on $FIS_d$ and $FIS_\alpha$, the two parameters $\omega_d$ and $\omega_\alpha$ can be easily obtained by (13), respectively.

The two weighted coefficients $\omega_d$ and $\omega_\alpha$ can largely affect the fused result simultaneously. Thus, the parameters $\omega_d$ and $\omega_\alpha$ need to be comprehensively considered as

$$\tilde{\omega}(p) = \omega_d(p) + \omega_\alpha(p), \tag{14}$$

where $\omega_d(p)$ and $\omega_\alpha(p)$ are the weighted coefficients of the data point at $p$-step.

Here, the minimum value $\omega_{\min}(p) = \min\{\tilde{\omega}_s(p), s=1,2,\cdots,L_n\}$ needs to be selected, with $L_n$ being the number of phase trajectories.

Finally, for the evolutionary trajectories $L_l$, the weighted coefficient $\omega_l(p)$ of the state point at $p$-step could be computed as

$$\omega_l(p) = \frac{\exp\{\gamma[\tilde{\omega}_l(p) - \omega_{\min}(p)]\}}{\sum_{s=1}^{L_n} \exp\{\gamma[\tilde{\omega}_s(p) - \omega_{\min}(p)]\}}, \tag{15}$$

where the parameter $\gamma$ is set to 1 in this paper.

Now the weighted coefficient of point on evolutionary trajectory can be calculated. By utilizing the weighted coeffcients, the NFDA algorithm can successfully complete the task of the data reduction. The steps of the approach are listed as follow.

1) Choose the proper embedding dimension $m_{max}$, delay time $\tau_{min}$, initial condition $X_F(0)$, and for each lead of 12-lead ECG signals, construct the vector $X_l$ as

$$X_l = (x(T), x(T+\tau_{min}), \cdots, x(T+(m_{max}-1)\tau_{min})) \in R^m, \quad (16)$$

where $T = 1, 2, \cdots, N-(m_{max}-1)\tau$ and $X_l$ is the evolutionary trajectory of the $l$th lead ECG signal on reconstructed trajectory. $X_F(0)$ is chosen as centroid of all the neighbor neighboring vectors $X_s(0) = (x(0), x(0+\tau_{min}), \cdots, x(0+(m_{max}-1)\tau_{min}))$, which is the evolutionary trajectory of the $s$th lead ECG signal at $p=0$.

Here the embedding dimension and delay time should be chosen as

$$m_{max} = \max\{m_s, s=1,2,\cdots,L_n\}, \quad (17)$$

$$\tau_{min} = \min\{\tau_s, s=1,2,\cdots,L_n\}, \quad (18)$$

2) For the $l$th evolutionary trajectory $X_l$, calculate the parameters $D(p)$, $D_r(p)$, $\alpha(p)$ and $\alpha_r(p)$ by (9)-(12), respectively. With two FISs, $\omega_d(p)$ and $\omega_\alpha(p)$ being properly estimated, the weighted coefficient $\omega_l(p)$ of the state $X_l(p)$ at the $p$-step can be computed by (14), (15).

3) Compute linear fitting parameters $a$ and $b$ by the least square equation

$$g(a,b) = \min_{a,b \in R}\{\sum_{s=1}^{L_n} \omega_s(p)[X_s(p+1)-ae-bX_s(p)]^2\}, \quad (19)$$

where $e = [1, \cdots, 1]_{m_{max}}^T$.

4) Via the linear prediction equation, the final fused state $X_F(p+1)$ is calculated as

$$X_F(p+1) = ae + bX_F(p). \quad (20)$$

In this section, the NFDA algorithm for 12-lead ECG signals is introduced. In the algorithm, the linear prediction equation is used to compute the fused state. With the qualitative characteristics of original signal, the weighted coefficient of each reconstructed trajectory is estimated through FIS properly. In the next section, the algorithm will be applied to 12-lead ECG signals and the performance of this algorithm will be further illustrated.

## 4. Application of NFDA in 12-lead ECG Signals

In order to estimate the quality condition of 12-lead ECG signals comprehensively and further improve the computational efficiency, the NFDA algorithm is proposed.

In this section, in order to assess the performance of NFDA, synthetic ECG signals and realistic ECG signals are applied in the experiments. In subsection 4.1, we evaluate the validity of NFDA by synthetic ECG signals. Then via three types of noises from the MIT-BIH Noise Stress Test Database (NSTDB) [26, 27], the noise tolerance of the algorithm will be analyzed in detail. In subsection 4.2, NFDA is executed on the database of PhysioNet/Computing in Cardiology Challenge 2011 [5] to further illustrate the performance of the algorithm. It is worth mentioning that, in this study, the False Neatest Neighbors (FNN) algorithm and the Average Displacement (AD) algorithm are adopted to determine the optimal embedding dimension $m_s$ and delay time $\tau_s$ of the *s*th lead ECG signal. The two methods can guarantee the objectivity and accuracy of the experiments to some extent.

### 4.1. Synthetic Signals Experiments

As realistic ECG signals are recorded in clinical environment, the signals would be contaminated inevitably by the noise and artifacts with different magnitudes. For solving this problem, synthetic ECG signals were widely applied in estimating the performance of algorithm. In [28], McSharry *et al*. proposed a dynamical model for generating synthetic ECG signals. Based on the idea of dynamical model, Sameni *et al*. [29] and Clifford *et al*. [30] developed an improved dynamical model which can generate 12-lead synthetic ECG signals. In the experiment, ideal VCG signals can be obtained via the model, as shown in Fig. 3 and it will be employed for testing the performance of NFDA.

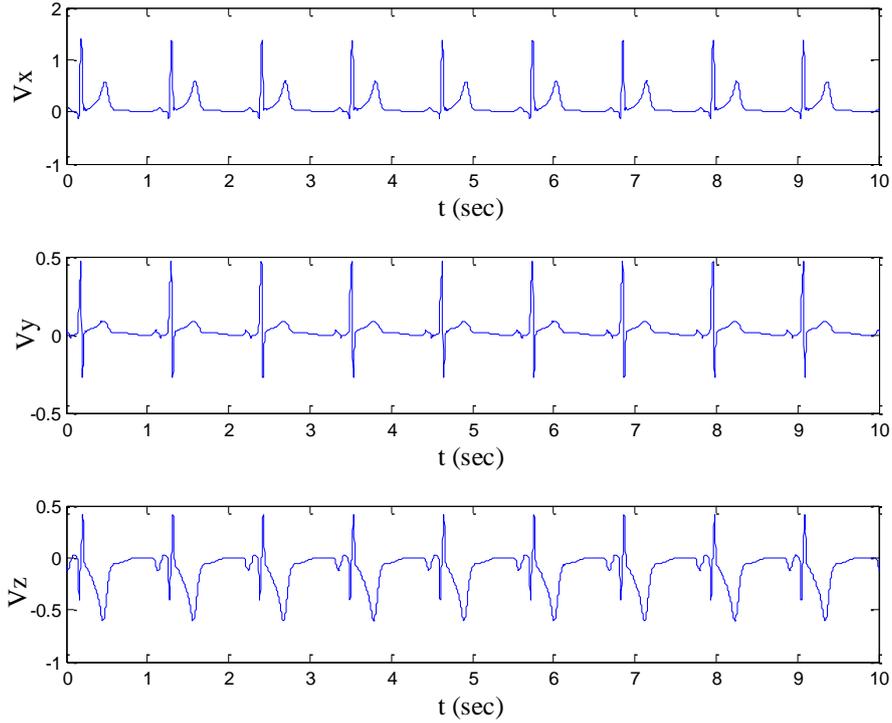

**Figure 3**. Synthetic ECG signals

In Fig. 4, the figures, (a)~(c), are the reconstructed trajectories of $V_x$, $V_y$ and $V_z$, respectively. From the morphology of reconstructed trajectory perspective, it is clear to see that there are needle-like features (Feature 1) on the three trajectories. Meanwhile, the longer closed trajectory (Feature 3) and the disorder feature of the closed trajectory within a small space (Feature 2) are shown distinctly in Figs. 4 (a)~(c). In Fig. 4 (a), the local trajectories of Features 2 & 3 essentially reflect the P wave and QRS complexes in ECG signal, respectively. The reconstructed trajectory of final fused result is shown in Fig. 4 (d), and evidently the three features of original VCG signals are well described by the fused trajectory.

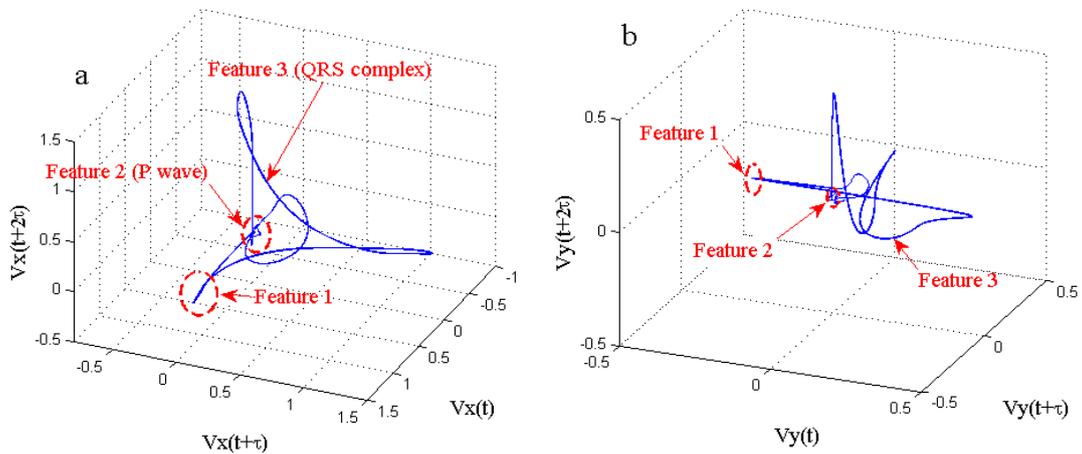

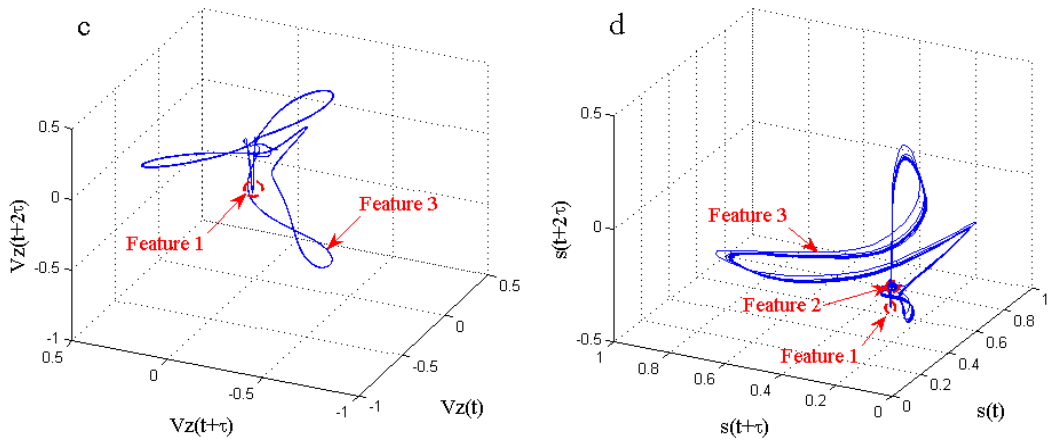

**Figure 4**. Reconstructed phase trajectories of synthetic ECG signals. (a)~(c) are reconstructed phase trajectories of $V_x$, $V_y$ and $V_z$, respectively. (d) is reconstructed phase trajectory of final fused result via NFDA.

Realistic ECG signals may be contaminated with the different types of the noises, e.g., baseline wander (BW), electrode movement (EM) and muscle artifact (MA), which cannot be easily removed by simple filter algorithms. Hence, the trajectory fusion problem of the noisy VCG signals will be discussed. To ensure the objectivity of experiment in this study, realistic noises are adopted from NSTDB and the three types of realistic noise, BW, EM and MA are shown in Fig. 5, respectively.

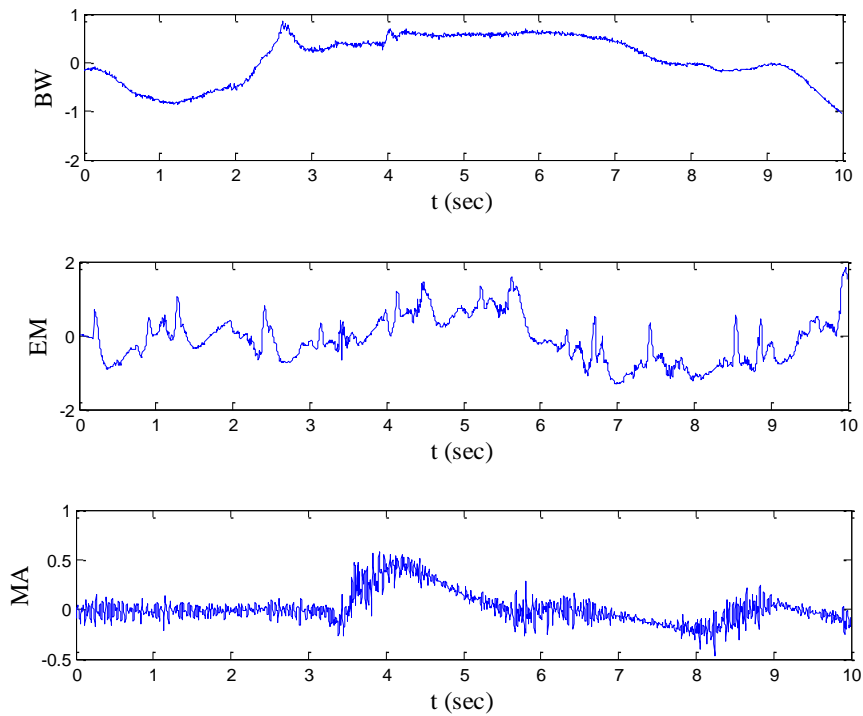

**Figure 5**. Realistic noise signals.

To further illustrate the validity of NFDA, the three types of noise, BW, EM and MA, are added to clean synthetic VCG signals, with different magnitudes of signal noise ratio (SNR). 12-lead ECG signals can be transformed to 3-lead VCG signals via the linear transformation. It means that the signal quality of 12-lead ECG can be inherited in some degree. In other words, if one lead signal in 12-lead ECG signals is contaminated by the noise, the quality characteristics of the lead signal will also be reflected in the VCG signals. In the experiment, the lead $V_x$ of VCG signals is randomly chosen, which is contaminated by the noise. The parameters of the SNR levels are summarized in Table 3 [12].

**Table 3:** SNR magnitudes for noise, BW, EM and MA.

|    | SNR levels (dB) | | | |
|----|----|---|----|-----|
| BW | 12 | 6 | 0  | -6  |
| EM | 6  | 0 | -6 | -12 |
| MA | 12 | 6 | 0  | -6  |

Here the lead $V_x$ is polluted by BW and the magnitudes of SNR are 12dB, 6dB, 0dB and -6dB, respectively. The reconstructed trajectories of noisy signals are shown in Figs. 6(a1), (b1), (c1) and (d1). In Fig. 6, these reconstructed trajectories of noisy signal $V_x$ present the divergent state clearly and the degree of divergence gradually increases with the SNR level decreasing correspondingly. Via NFDA, the noisy VCG signals are fused to some extent. Similarly, the divergent states of reconstructed trajectories are reflected in Figs. (a2), (b2), (c2) and (d2).

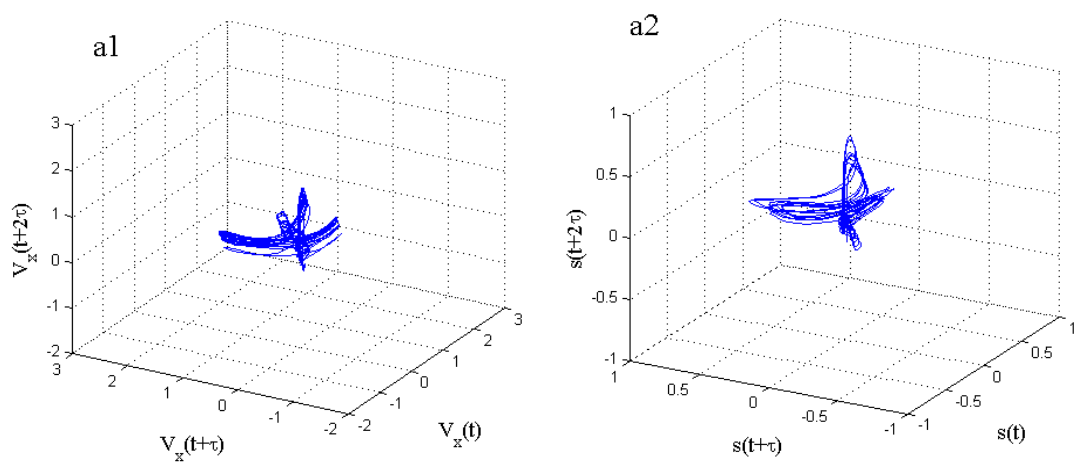

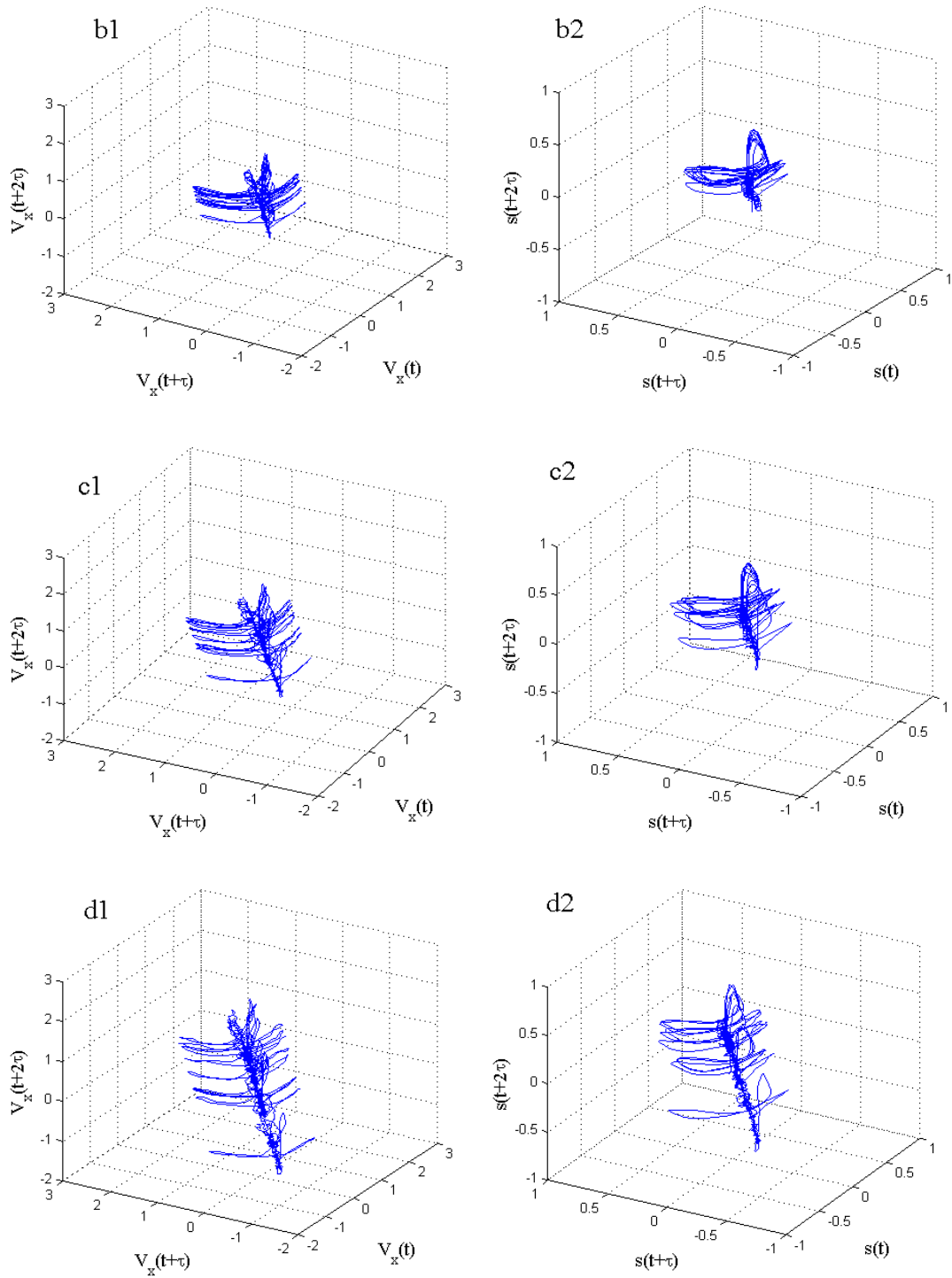

**Figure 6**. Reconstructed trajectories of noisy $V_x$ signal and fused results. (a1)~(d1) are reconstructed trajectories of $V_x$ signal contaminated by BW with SNR being 12dB, 6dB, 0dB and -6dB, respectively. (a2)~(d2) are reconstructed trajectories of corresponding fused results.

In this section, the clean $V_x$ signal is polluted by the noise of EM and MA with different magnitudes of SNR and the reconstructed trajectories are shown in Fig. 7 and Fig. 8, respectively.

In the two figures, the relations between noisy signals and fused results are highly consistent with the relation reflected in Fig. 6.

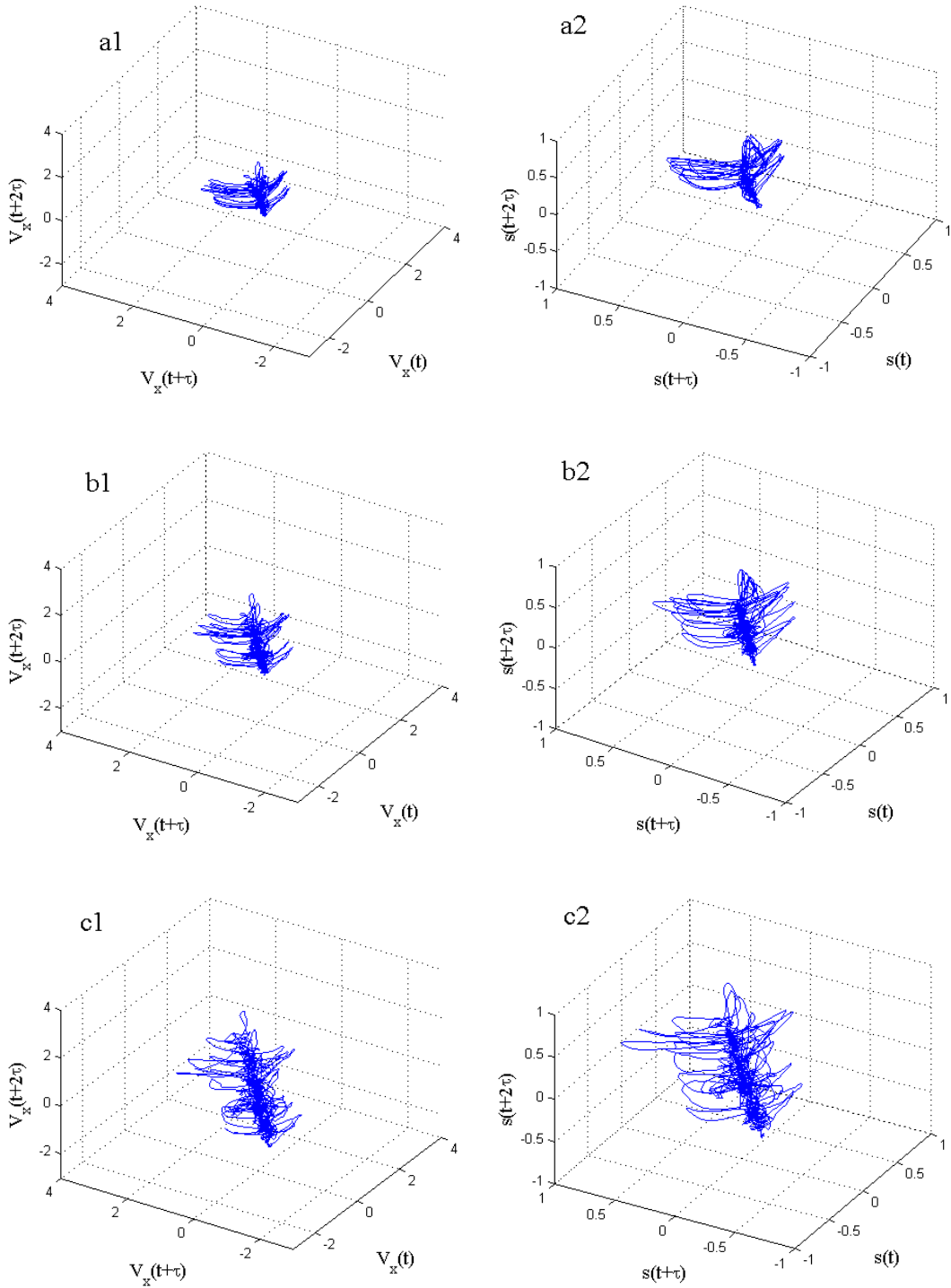

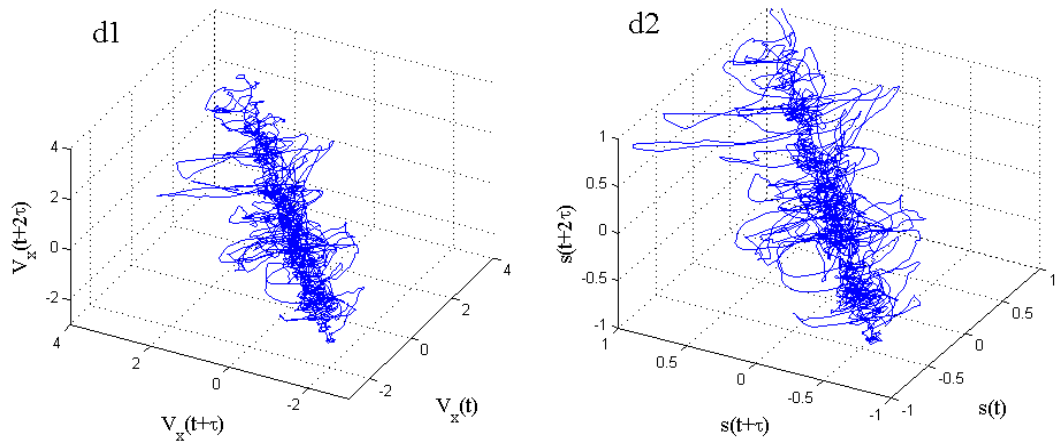

**Figure 7**. Reconstructed trajectories of noisy $V_x$ signal and fused results. (a1)~(d1) are reconstructed trajectories of $V_x$ signal contaminated by EM with SNR being 6dB, 0dB, -6dB and -12dB, respectively. (a2)~(d2) are reconstructed trajectories of corresponding fused results.

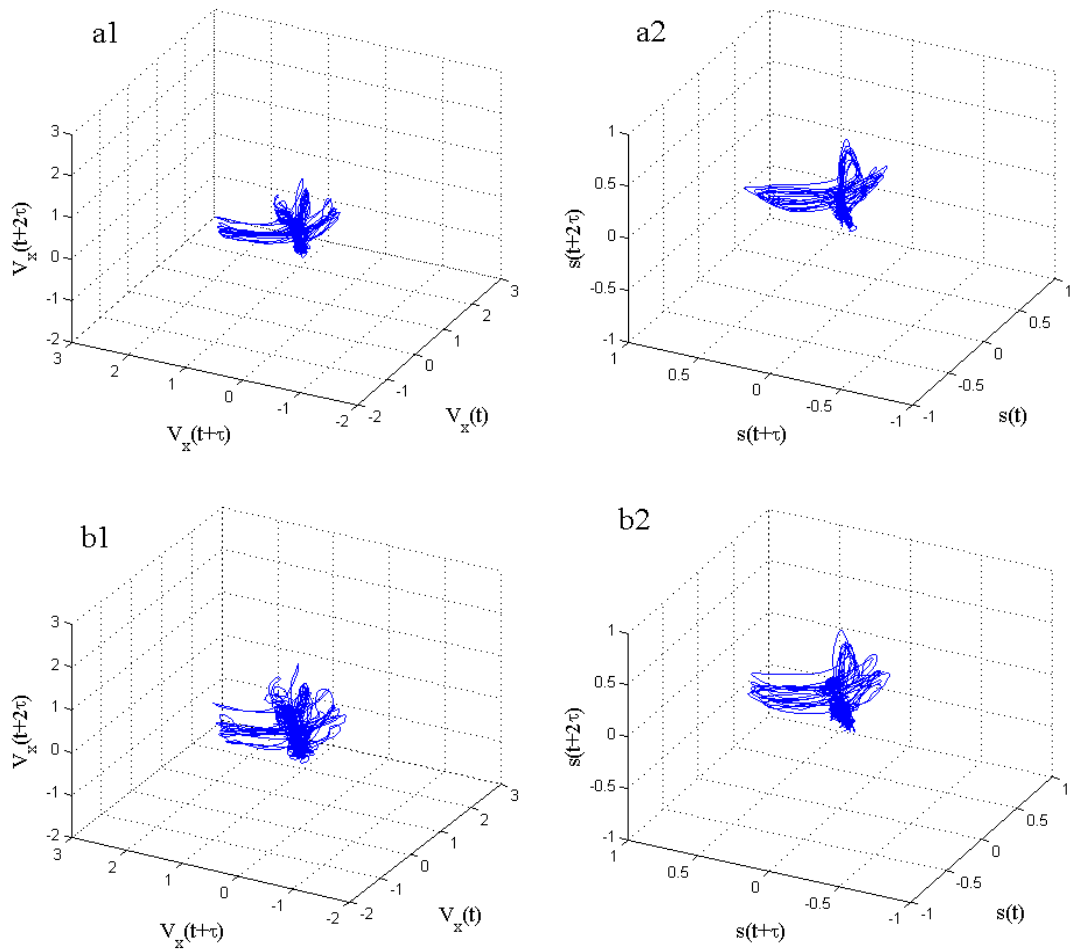

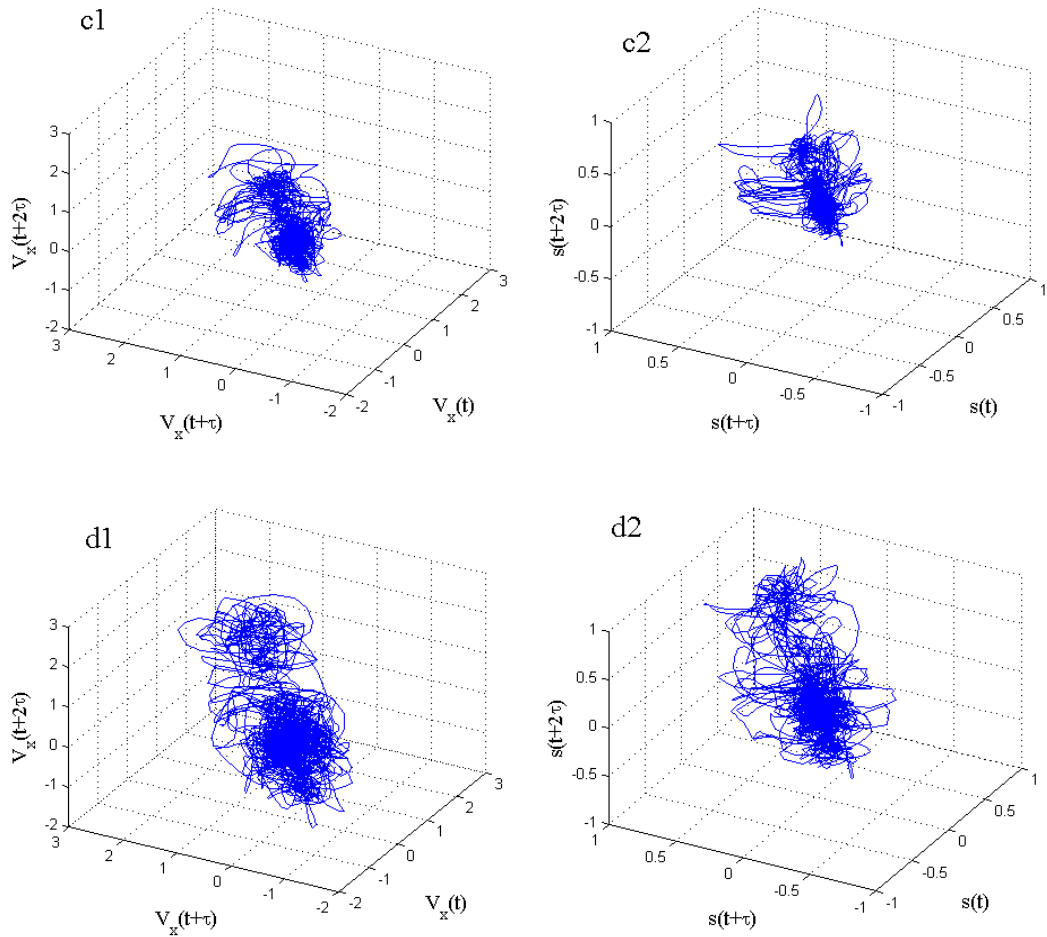

**Figure 8.** Reconstructed trajectories of noisy $V_x$ signal and fused results. (a1)~(d1) are reconstructed trajectories of $V_x$ signal contaminated by MA with SNR being 12dB, 6dB, 0dB and -6dB, respectively. (a2)~(d2) are reconstructed trajectories of corresponding fused results.

From Fig. 8, we can find that Figs. 8 (a1)~(d1) are the reconstructed trajectories of noisy $V_x$. The disordering degree of the reconstructed trajectory increases progressively with the magnitude of SNR decreasing. On the other hand, Figs. 8 (a2)~(d2) are the reconstructed trajectories of the fused signals. Analogously, there is a negative correlation between the disordering degree of reconstructed trajectory and the SNR level of noisy $V_x$.

With the experimental results being comprehensively analyzed under different conditions, it suggests that with the NFDA algorithm being employed, the reconstructed trajectories of fused results can effectively describe the quality characteristics of noisy synthetic ECG signals. In order to test the performance of our method adequately, some realistic ECG signals will be applied in subsection 4.2.

## 4.2. Realistic Signals Experiments

As an important database, the PhysioNet/Computing in Cardiology Challenge 2011 has been widely used for testing the ECG quality assessment algorithms. In the database, standard 12-lead ECG signal is sampled at 500Hz and recorded for 10 seconds. There are 1000 12-lead ECG records to be employed as train set (Set A) and the signal quality was quantified by a group of annotators professional in ECG analysis. According to the quantitative result of ECG signal, 773 ECG records in Set A are acceptable, 225 signals are unacceptable and the remaining are indeterminate.

In this subsection, four sets of realistic 12-lead ECG signals are randomly selected from set A for assessing the performance of NFDA. Thereinto, the quality of No. 1027085 and No. 1075113 are acceptable and the quality of No. 1063069 and No. 1003574 are unacceptable.

It is important to note that the length of data in set A is 10 seconds. In clinical setting, ECG signals might be contaminated by the uncertain noise in a certain time period. In other words, for a long-term ECG signal, sometimes only parts of the ECG signal quality are unacceptable with the remaining being acceptable. Therefore, the long-term ECG signal needs to be divided into a number of short-term ECG signals and the signal quality of them could be estimated individually.

In order to analyze the ECG signal quality, the realistic ECG signals need to be pre-processed before quality assessment. Firstly, each lead of the 12-lead ECG signals should be examined for the constant signal detection. If some constant signals are contained in the ECG signals, then the realistic signal need not be further processed and it can be identified as being unacceptable. Otherwise, the 12-lead ECG signals will be transformed into VCG signals by the inverse Dower transformation matrix [31] by the following equation:

$$VCG = D_{inv} \times ECG, \qquad (25)$$

where $D_{inv}$ is given by

$$D_{inv} = \begin{bmatrix} -0.172 & -0.074 & 0.122 & 0.231 & 0.239 & 0.194 & 0.156 & -0.010 \\ 0.057 & -0.019 & -0.106 & -0.022 & 0.041 & 0.048 & -0.227 & 0.887 \\ -0.229 & -0.310 & -0.246 & -0.063 & 0.055 & 0.108 & 0.022 & 0.102 \end{bmatrix}. \qquad (26)$$

By utilizing the transformation, the quality characteristics of original signals can be completely inherited by VCG signals. Then the VCG signals will be processed by NFDA.

*Remark 1*: There is redundant information in the 12-lead ECG signals, and the leads III, avR, avL and avF can be computed from other leads. In other words, the leads III, avR, avL and avF are redundant leads in Dower transformation. Thus, there are eight leads being used in the Dower transformation algorithm.

By utilizing NFDA, the VCG signals and reconstructed trajectory of fused result for the signals of No. 1027085 and No. 1075113 are shown in Figs. 9 (a) and (b). As the qualities of the two signals are acceptable, the VCG signals demonstrate the periodic changes and possess significant physiological meaning. Further, the reconstructed trajectories of fused results imply the regular evolutionary characteristics. On the contrary, the other two realistic signals are unacceptable. From the Figs. 9(c) and (d), the VCG signals and fused results indicate discursive and unsystematic evolutionary characteristics.

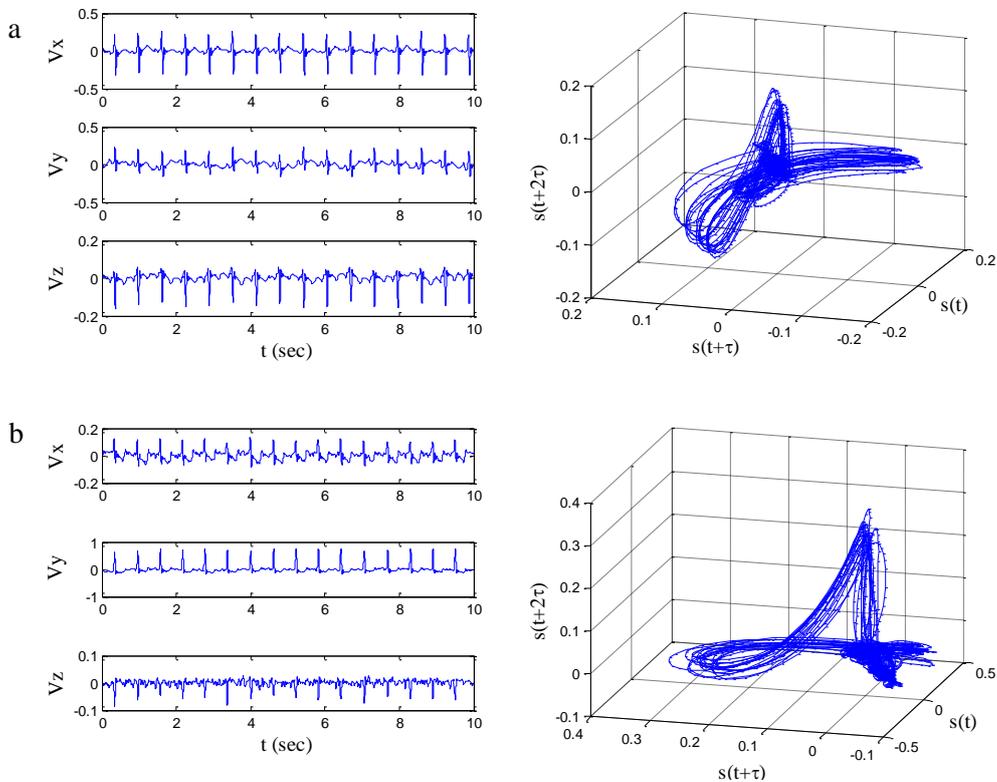

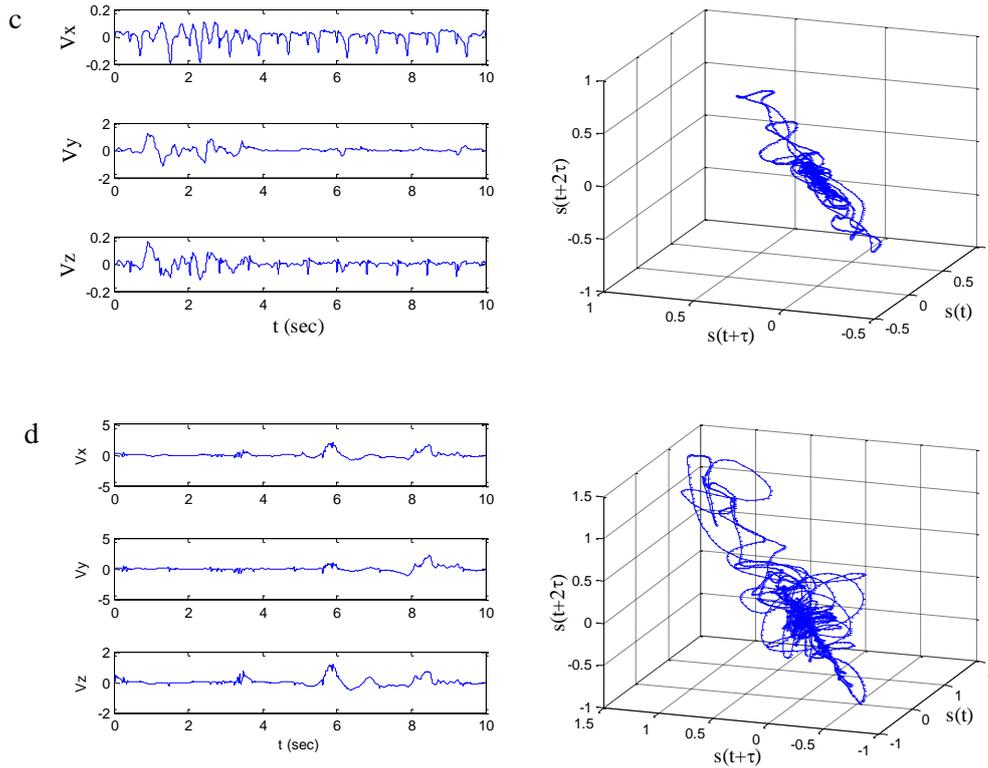

**Figure 9**. 3-lead VCG signals by realistic 12-lead ECG signals transformation and reconstructed trajectories of fused results. (a)~(d) are results of the signal of No.1027085, No.1075113, No.1027085 and No.1075113, respectively.

The NFDA algorithm stresses that the quality characteristics of ECG signal with higher frequency and larger fluctuation should be inherited by the fused trajectory. It means that the characteristic information of some lead ECG signals with visible time-domain properties are well inherited. Whereas time domain properties of other leads signal are diminished inevitably in the fused trajectory.

In this subsection, NFDA is evaluated by realistic 12-lead ECG signals. Experimental results indicate that the fused trajectory can effectively inherit the quality characteristics of 12-lead ECG signals.

## 5. Conclusion

In this paper, the NFDA algorithm is proposed, which utilizes the idea of the LWLPA algorithm to fuse 12-lead ECG signals. Meanwhile, two fuzzy inference systems are designed for effectively inheriting the characteristics of original signals.

In this study, Synthetic ECG signals, noisy synthetic ECG signals and realistic ECG signals are employed to test the validity of the algorithm. Due to the limitation of paper length, two 12-lead ECG signals are adopted randomly from Set A of PhysioNet/Computing in Cardiology Challenge 2011 which contains 773 acceptable qualities of ECG records. Analogously, two 12-lead ECG signals are adopted randomly from the data set which are tagged as unacceptable quality. By the analysis of the remaining data in Set A, the quality characteristics of ECG signals can be exhibited by the reconstructed trajectories of the fused signals clearly. The experimental results indicate that the NFDA algorithm could effectively compress the 12-lead ECG signals and well fuse the quality characteristics of the original signal.

There are still many problems awaiting us to offer solutions. If the fused signal need to be analyzed further, how to obtain the quantified characteristic parameters is a crucial problem in the quality estimation of ECG signal, although the quality characteristics of the fused signal can be observed easily. The recurrence quantification analysis (RQA) method particularly suits to handle the biological signals. Hence, RQA will be employed to quantificationally extract the quality characteristics of the fused signal in the further research. Additionally, how to design an optimized FIS is also needs to be dealt with as the future work.

## 6. Acknowledgments

This work is supported by National Natural Science Foundation (NNSF) of China (Grant 61374054), Natural Science Foundation of Gansu Province (Grant 17JR5RA278) and Fundamental Research Funds for the Central Universities (Grants 3192015007, 31920170015, 31920170141 & 31920160003).

## 7. Conflict of Interest

The authors declare that there is no conflict of interest regarding the publication of this paper.